\PassOptionsToPackage{unicode}{hyperref}
\PassOptionsToPackage{hyphens}{url}
\PassOptionsToPackage{dvipsnames,svgnames,x11names}{xcolor}
\documentclass[
  12pt]{article}

\usepackage{amsmath,amssymb}
\usepackage{iftex}
\ifPDFTeX
  \usepackage[T1]{fontenc}
  \usepackage[utf8]{inputenc}
  \usepackage{textcomp} % provide euro and other symbols
\else % if luatex or xetex
  \usepackage{unicode-math}
  \defaultfontfeatures{Scale=MatchLowercase}
  \defaultfontfeatures[\rmfamily]{Ligatures=TeX,Scale=1}
\fi
\usepackage{lmodern}
\ifPDFTeX\else  
\fi
\IfFileExists{upquote.sty}{\usepackage{upquote}}{}
\IfFileExists{microtype.sty}{% use microtype if available
  \usepackage[]{microtype}
  \UseMicrotypeSet[protrusion]{basicmath} % disable protrusion for tt fonts
}{}
\makeatletter
\@ifundefined{KOMAClassName}{% if non-KOMA class
  \IfFileExists{parskip.sty}{%
    \usepackage{parskip}
  }{% else
    \setlength{\parindent}{0pt}
    \setlength{\parskip}{6pt plus 2pt minus 1pt}}
}{% if KOMA class
  \KOMAoptions{parskip=half}}
\makeatother
\usepackage{xcolor}
\makeatletter
\ifx\paragraph\undefined\else
  \let\oldparagraph\paragraph
  \renewcommand{\paragraph}{
    \@ifstar
      \xxxParagraphStar
      \xxxParagraphNoStar
  }
  \newcommand{\xxxParagraphStar}[1]{\oldparagraph*{#1}\mbox{}}
  \newcommand{\xxxParagraphNoStar}[1]{\oldparagraph{#1}\mbox{}}
\fi
\ifx\subparagraph\undefined\else
  \let\oldsubparagraph\subparagraph
  \renewcommand{\subparagraph}{
    \@ifstar
      \xxxSubParagraphStar
      \xxxSubParagraphNoStar
  }
  \newcommand{\xxxSubParagraphStar}[1]{\oldsubparagraph*{#1}\mbox{}}
  \newcommand{\xxxSubParagraphNoStar}[1]{\oldsubparagraph{#1}\mbox{}}
\fi
\makeatother

\usepackage{longtable,booktabs,array}
\usepackage{calc} % for calculating minipage widths
\usepackage{etoolbox}
\makeatletter
\patchcmd\longtable{\par}{\if@noskipsec\mbox{}\fi\par}{}{}
\makeatother
\IfFileExists{footnotehyper.sty}{\usepackage{footnotehyper}}{\usepackage{footnote}}
\makesavenoteenv{longtable}
\usepackage{graphicx}
\makeatletter
\def\maxwidth{\ifdim\Gin@nat@width>\linewidth\linewidth\else\Gin@nat@width\fi}
\def\maxheight{\ifdim\Gin@nat@height>\textheight\textheight\else\Gin@nat@height\fi}
\makeatother
\setkeys{Gin}{width=\maxwidth,height=\maxheight,keepaspectratio}
\makeatletter
\def\fps@figure{htbp}
\makeatother

\makeatletter
\@ifpackageloaded{caption}{}{\usepackage{caption}}
\AtBeginDocument{%
\ifdefined\contentsname
  \renewcommand*\contentsname{Table of contents}
\else
  \newcommand\contentsname{Table of contents}
\fi
\ifdefined\listfigurename
  \renewcommand*\listfigurename{List of Figures}
\else
  \newcommand\listfigurename{List of Figures}
\fi
\ifdefined\listtablename
  \renewcommand*\listtablename{List of Tables}
\else
  \newcommand\listtablename{List of Tables}
\fi
\ifdefined\figurename
  \renewcommand*\figurename{Figure}
\else
  \newcommand\figurename{Figure}
\fi
\ifdefined\tablename
  \renewcommand*\tablename{Table}
\else
  \newcommand\tablename{Table}
\fi
}
\@ifpackageloaded{float}{}{\usepackage{float}}
\floatstyle{ruled}
\@ifundefined{c@chapter}{\newfloat{codelisting}{h}{lop}}{\newfloat{codelisting}{h}{lop}[chapter]}
\floatname{codelisting}{Listing}

\makeatother
\makeatletter
\@ifpackageloaded{caption}{}{\usepackage{caption}}
\@ifpackageloaded{subcaption}{}{\usepackage{subcaption}}
\makeatother

\ifLuaTeX
  \usepackage{selnolig}  % disable illegal ligatures
\fi
\usepackage[]{natbib}
\usepackage{bookmark}

\IfFileExists{xurl.sty}{\usepackage{xurl}}{} % add URL line breaks if available
\hypersetup{
  pdftitle={The Evolution and Interpretation of ``Statistical Purposes''},
  pdfauthor={Michael B. Hawes; John L. Eltinge; Paul S. Marck; Danielle C. Neiman; and Sallie Ann Keller},
  pdfkeywords={data quality; disclosure avoidance; legal and regulatory enforcement; privacy protection; public benefit; public trust},
  colorlinks=true,
  linkcolor={blue},
  filecolor={Maroon},
  citecolor={Blue},
  urlcolor={Blue},
  pdfcreator={LaTeX via pandoc}}

\newcommand{\anon}{1}

\begin{document}

\def\spacingset#1{\renewcommand{\baselinestretch}%
{#1}\small\normalsize} \spacingset{1}

%%%%%%%%%%%%%%%%%%%%%%%%%%%%%%%%%%%%%%%%%%%%%%%%%%%%%%%%%%%%%%%%%%%%%%%%%%%%%%

\if1\anon
{
  \title{\bf The Evolution and Interpretation of ``Statistical Purposes''}
  \author{Michael B. Hawes\thanks{
    Corresponding Author (michael.b.hawes@census.gov);\\ Earlier versions of the material included in this paper were presented by Sallie Ann Keller at the 2023 Federal Committee on Statistical Methodology Research and Policy Conference, by Danielle Neiman at the 2024 European Conference on Quality in Official Statistics in Estoril, Portugal, and by Michael Hawes and Sallie Ann Keller in 2026 for the American Statistical Association and the Social Data Science Center. Any opinions or viewpoints expressed in this paper are the authors' own and not the opinions or viewpoints of the U.S. Census Bureau.}\hspace{.2cm}\\
    \bigskip    U.S. Census Bureau\\
    John L. Eltinge\\
    \bigskip    U.S. Census Bureau, retired\\
    Paul S. Marck\\
    \bigskip    U.S. Census Bureau\\
    Danielle C. Neiman\\
    \bigskip    U.S. Census Bureau\\
    \bigskip    and \\
    Sallie Ann Keller \\
    University of Virginia; formerly U.S. Census Bureau}
  \maketitle
} \fi

\if0\anon
{
  \bigskip
  \bigskip
  \bigskip
  \begin{center}
    {\LARGE\bf The Evolution and Interpretation of ``Statistical Purposes''}
\end{center}
  \medskip
} \fi

\bigskip
\begin{abstract}
National Statistical Organizations (NSOs) and other groups often use the term “for statistical purposes only” in communication with prospective respondents, data users and other stakeholders. This term also provides an important anchor for many NSO decisions on operations and ethics. Although public communication often omits a clear operational definition of this term, this paper will show that reviews of underlying laws and policies identify two predominant criteria: (1) production of statistical information about relatively large population aggregates, with the intention of creating a general public benefit; and (2) protection of the confidentiality of data collected about data subjects and a related prohibition against the use of information provided by or about data subjects for legal or regulatory action against those individuals or organizations. We then explore how this term exists, and is often interpreted, within a much broader landscape of legal requirements and of scientific and professional codes of practice, and provide a broader definition for consideration based on these and other ethical frameworks. This paper closes by highlighting several areas that warrant further discussion to better position NSOs to navigate these challenges in the future.\end{abstract}

\noindent%
\vfill

\newpage
\spacingset{1.8} % DON'T change the spacing!

\section{Introduction}
\label{sec:intro}

When designing survey instruments and their accompanying communications materials, National Statistical Organizations (NSOs) and other statistical organizations   often prioritize brevity in communicating the purpose of data collection to their subjects. This may result in unanticipated  ambiguity in important messages that need to be conveyed. One common example of this is the statement that information provided will be used only for ``statistical purposes” (\cite{USCB:privacyprinciples}, \cite{BLS:mxp}, \cite{NCHS:NEHRS}). In spite of the ubiquity of this phrase, however, the term largely lacks a clear and succinct definition, and research indicates that it is poorly understood by the public \citep{landreth2008report}. Beyond the point of data collection, the term is also widely used in statute (44 US Code 3572(a)(3)), regulation (26 CFR 301.6103(j)(1)-1), and agency policies and standards \citep{NCES:standards},  and it routinely guides NSOs in their evaluation and interpretation of the activities that they can or should do. In this context, the phrase serves as a critical legal and ethical standard for NSO staff to follow in their day-to-day activities and decision making. But, to the extent that definitions or explanations of this phrase are provided, they frequently lack components of the concept that agency staff typically assume to be included and do not include other elements that could serve as more comprehensive guardrails for ``acceptable'' data use for NSO staff to follow. 

The phrase ``only for statistical purposes” is a useful external message and internal data use standard only to the extent that it accurately captures and conveys its intended meaning. Ambiguity or misunderstanding with regard to the term’s meaning not only weaken its use in communications to data subjects, but can also obscure the boundaries of potentially acceptable or unacceptable NSO activities. In this paper, we examine the historical origins of NSOs' use of the term ``statistical purposes,” documenting how the phrase has been routinely used throughout the U.S. federal statistical system. We then discuss and document the ambiguities and limitations inherent to these commonly used definitions and explanations. We examine the unstated elements that are typically assumed to be part of the definition, as well as those elements that are notably absent from the definitions but that perhaps merit inclusion if the term is to be of broader use as a more comprehensive ethical and managerial standard for NSOs to follow. With these ambiguities and limitations in mind, we identify some core elements in common use of the term in the context of various frameworks within which U.S. federal statistical agencies operate. We also compare use of the term by U.S. statistical agencies to select examples from across the international statistical community. We then propose some factors that NSOs may want to consider including in their future interpretations of what ``statistical purposes'' ought to mean in the context of official statistics and provide a broader, applied definition of the term for agencies to consider. We conclude with a series of suggestions and ideas for future exploration and research on this important topic.

\section{Historical Origins and Current Definitions of `Statistical Purposes' in the United States}
\label{sec:origins}
The U.S. federal statistical system, a decentralized network of statistical agencies and units with over 100  statistical programs across government \citep{OMB:stats}, collects and disseminates detailed statistical information covering the entire breadth of the nation's social and economic activities and characteristics. The statistical products these agencies produce serve as critical inputs for evidence-based policymaking by federal, state, tribal, and local officials, for high-value decision-making by the commercial and financial sectors, and for important scientific research in the fields of economics, public health, and demography, to name a few. The Census Bureau’s decennial census of population and housing, for example, is used to apportion seats in the U.S. House of Representatives, to draw legislative and voting district boundaries, and to allocate over \$2.8 trillion in federal funds annually \citep{VillaRoss2023FederalFundsFY2021}. In short, federal statistics are an important public good provided by the federal government, the benefits of which impact every corner of the nation.

\subsection{Early 20th Century Origins and the Second World War}

The phrase ‘statistical purposes’ has been used extensively across the entirety of the U.S. federal statistical system for decades to mean something beyond the mere practice of statistics as a discipline involving ``the systematic collection and arrangement of numerical facts or data of any kind...'' \citep{oed}. In fact, the meaning of “statistical purposes” as used within the federal statistical system starts with, but goes well beyond, a central premise of the public value and benefit of official statistics.\footnote{We distinguish ``public benefit” in this context more broadly than the stricter concept of a “public good” (e.g., a non-excludable and non-rivalrous product or service) as used in economics and public policy \citep{reiss2021public}. While official statistics often meet this formal definition of public good, their public benefit exists even in those situations when they do not. See also \cite{nickson2024statistics}}  The outlines of this benefit-focused use appear as early as 1910 with President Taft's Proclamation about the 1910 Census, which declared the census to be ``of the utmost importance to the interests of all people in the United States,'' indicating its ``sole purpose'' to ``secure general statistical information'' and promising that ``The census has nothing to do with taxation, with army or jury service, with the compulsion of school attendance, with the regulation of immigration, or with the enforcement of any national, state, or local law or ordinance, nor can any person be harmed in any way by furnishing the information required'' \citep{taft}. While statistical \textit{methods} are frequently used in ways that can directly harm an individual (e.g., to predict violent crime recidivism in criminal sentencing, or to flag tax returns for subsequent auditing), the statistical \textit{purpose} of the census, as proclaimed by President Taft, clearly hinged on the public benefit of the product and the corresponding need to encourage cooperation through promises that the ``general statistical information'' would not be used to harm the individual. This implicit linkage of public benefit and constrained use of the information to produce general, non-identifiable statistical information would continue over the subsequent century as the term ``statistical purposes'' came into more common use.

In spite of President Taft's pronouncement that an individual's identifiable census information would not be used against them, the 1910s and 1920s saw frequent use of census data for legal cases relating to draft evasion, as well as immigration and deportation proceedings \citep{gatewood2001monograph}. The Census Bureau even compiled lists of respondent names by literacy status for state and local officials in support of literacy campaigns (p.13). This laissez-faire attitude about the use of identifiable information soon ended, at least temporarily, with the passage of the 1929 Census Act, which provides what may be the first extant use of our term of interest, specifying that ``the information furnished under the provisions of this Act shall be used \textit{only for the statistical purposes} for which it is supplied. No publication shall be made...whereby the data furnished by any particular establishment or individual can be identified, nor shall the Director of the Census permit anyone other than the sworn employees of the Census Office to examine the individual reports'' (P.L. 71-12 \S 11, emphasis added). Though section 18 of the same act gave the Director of the Census the seemingly contradictory authority, at his discretion, to disclose individual, identifiable records ``Provided, however, that in no case shall information furnished under the authority of this Act be used to the detriment of the person or persons to whom such information relates.'' This new, though certainly not ironclad, commitment to purpose specification and limitation was quickly put to the test the following year when Attorney General William D. Mitchell affirmed that a formal request for identifiable Census information from the federal Women's Bureau (and subsequent requests from law enforcement) were now legally prohibited \citep[p.14]{gatewood2001monograph}. 
  
Soon, however, the exigencies of another world war first strained, then fully shattered this emerging idea that the \textit{statistical purposes} for which these data are collected preclude their use in a manner that could directly harm the individual data subjects. In early 1942, the Census Bureau assigned a statistician to assist the war agencies to provide on-demand census tract- and block-level aggregate tabulations of Japanese Americans living in California for use in identifying where to focus efforts for rounding up these individuals as part of the internment effort \citep[p.14-15]{gatewood2001monograph}. While it does not appear that any person-level identifiable information from the census was directly shared with the war agencies or law enforcement as part of this effort, the significant \textit{group harm} resulting from this use of the data is clearly evident. Even this line in the sand was quickly abandoned when the Second War Powers Act (1942) formally repealed the confidentiality of census data (P.L. 507 \S 1402), allowing federal agencies to request copies of individual census records. Most notable among the subsequent disclosures of identifiable 1940 Census data was a list provided by the Census Bureau to the U.S. Secret Service containing the names, addresses, occupations, and citizenship status of all Japanese Americans residing in the Washington, D.C. metropolitan area \citep{seltzer2007census}.    

\subsection{Second Half of the Twentieth Century: Increasing Emphasis on Privacy Protections}

By the late 1940s most of the Second War Powers Act was repealed, returning the confidentiality guarantee of the decennial census. These protections were then further strengthened with the 1954 codification of Title 13 into the U.S. Code (13 U.S. Code, 1954). In this codification we again see the phrase “statistical purposes” as part of these protections, ``...for any purpose other than the statistical purposes for which it is supplied...'' (13 U.S. Code \S 9(a)(1), but, as in the 1929 Census Act, the term remained undefined, and discretionary (\S 8(a)).

The 1970s saw a new federal focus on privacy rights and protections. Most notable among these, for this discussion, were the passage of the Privacy Act of 1974, which imposed new, government-wide restrictions on the disclosure of identifiable information from systems of records and the subsequent 1976 amendment to Title 13 curtailing the Census Bureau director's discretion to release identifiable information under \S 8(a) \citep{events}, narrowing and clarifying the implied interpretation of statistical purpose. In addition, the Privacy Protection Study Commission of 1977 introduced the concept and importance of \textit{functional separation} of federal statistical activities from federal administrative or enforcement actions. In doing so, the commission’s report uses the term “research or statistical purpose.” Specifically, their definition of functional separation entails “separating the use of information about an individual for a research or statistical purpose from its use in arriving at an administrative or other decision about that individual” \citep[Chapter 15]{united1977personal}.

\subsection{Entering the Twenty-First Century: More Formal Codification}

By the 1990s, declining trust in government and growing concerns about data privacy in the nascent internet age, prompted Katherine Wallman, the Chief Statistician of the United States at the Office of Management and Budget, to initiate work leading to an order enhancing the purpose specification and limitations of federal statistical collections (62 Fed. Reg. 35,044). This order at last provided a formal federal definition of statistical purpose, as ``...the description, estimation or analysis by the Federal Government of information concerning persons, the economy, society, or the natural environment (or relevant groups or components thereof) without regard to the identities of specific persons...[and] excludes many other activities or functions for which information is used in identifiable form, such as determining whether a person is eligible for a license, privilege, right, grant, or benefit...or whether a person's conduct was or is in accordance with law...'' (\S1(i)). The 2002 passage of the Confidential Information Protection and Statistical Efficiency Act (CIPSEA, P.L. 107-347, 2002), put this language into statute, nearly verbatim, via the Act's definitions of \textit{statistical purpose} \S502(7) and \textit{nonstatistical purpose} (\S502(5)). This statutory definition was later reaffirmed, unchanged, when CIPSEA was amended with the passage of the Foundations for Evidence-based Policymaking Act of 2018 (known colloquially as the ``Evidence Act''), which President Donald Trump signed into law in January 2019. As noted at the end of Table 1, the Evidence Act also included additional definitions that complement and provide further context for agencies' interpretation of ``statistical purposes.''

Through the evolution of the term \textit{statistical purpose} over the 109 years between President Taft's 1910 proclamation and President Trump's signature of the Evidence Act in 2019, we see clear commonality relating to confidentiality, constrained use, and the prevention of harm to the individual. CIPSEA's definitions of statistical and nonstatistical purpose together convey a strong notion of confidentiality: ``the description, estimation or analysis of the characteristics of groups, without identifying the individuals or organizations that comprise such groups'' (\S502(5)). They also convey a strong commitment not to \textit{use} identifiable information about individuals in a manner that can harm those individuals: ``...use of data in identifiable form for any purpose that is not a statistical purpose, including any administrative, regulatory, law enforcement, adjudicatory, or other purpose that affects the rights, privileges, or benefits of a particular identifiable respondent'' (\S502(7)). Legal definitions of ``statistical purposes,” such as in CIPSEA focus on permissible uses of statistical data, such as uses that describe or estimate the characteristics of groups without revealing or identifying the data subjects that belong to those groups. As such, this restriction on permissible uses is a central component of the interpretation of the definition. However, embedded within that definition, and in the broader context in which it is typically used, is another dimension of restrictions: limits on who can access the identifiable information about data subjects that are the necessary precursors to the production of the statistics. These two dimensions are intended to work in tandem to help reassure data subjects that they will not be adversely impacted by participating in a survey: the data can only be used to produce statistics about groups, and they can only be accessed by the individuals who will be contributing to the development and production of those statistics. Taken together, the access and use restrictions inherent to the statutory definition of statistical purpose serve to cement the statutory intention of providing functional separation of statistical activities - the insulation of the production of official statistics from the broader administrative and enforcement activities of government. 

Notably absent from the formal CIPSEA definitions, however, is the explicit exhortation that served as the underlying motivation for Taft's pronouncement: the notion that this fundamental statistical purpose is ultimately in pursuit of a notable public benefit. And though the 1997 confidentiality order that served as the subsequent impetus for CIPSEA and the Evidence Act was explicitly motivated by the need to improve \textit{trust}, that conception of trust was explicitly about confidentiality and protection from harm. As we will see in Section \ref{sec:trust}, trust in official statistics is a bigger issue that our conception of \textit{statistical purpose} may need to also address.

\section{Declining Trust, Growing Demand for Objective and Reliable Evidence}
\label{sec:trust}

In 1964, fully 77\% of Americans trusted the federal government ``to do what is right just about always/most of the time'' according to the Pew Research Center. Over the subsequent decades, that trust did not just erode, it cratered, such that by 2024, a mere 23\% of the population felt similarly \citep{pew}. Though trust in federal statistical agencies is harder to measure \citep{hunter2019trust}, the precipitous decline in response rates to federal surveys over the past two decades suggests that public trust in official statistics may be likewise waning \citep{czajka2016background, acs}. While broader societal attitudes about privacy and confidentiality concerns are likely a partial driver of these trends \citep{madden2015americans}, stronger messaging about confidentiality protections in the context of statistical collections seems to only heighten, rather than alleviate, respondents' aggregate levels of distrust \citep[p. 259]{singer1992confidentiality}. In this environment, the term \textit{statistical purposes} is especially problematic. 

In fact, cognitive testing has indicated that respondents may misunderstand the phrase  \citep{gerber2000privacy, landreth2001sipp}. While some respondents may understand what \textit{statistics} are, they may not associate this term with any degree of confidentiality protection \citep[p.19-20]{landreth2008report}. Ironically, even when faced with messaging about how their data would be confidential and only used for statistical purposes, many respondents in the Landreth, et al. study (2008, pp. 24-25) reported the perception that their information would be shared with other government agencies.

Faced with the growing trust deficit in government institutions, and the parallel decline in response rates to federal statistical collections, the continued ability of federal statistics to meet the demand to support evidence-based decision-making (the public benefit) hinges on the ability of statistical agencies to protect the confidentiality of respondent information (to encourage participation) but also on the overall objectivity, integrity, and representativeness of the data being collected and the statistics being produced (to instill trust not just in the accuracy and validity of the statistics being produced, but also indirectly in the reliability of the decisions and conclusions that are based upon them). Are the statistics accurate? Are they unbiased? And do they support the broader needs of the nation, including small but important and under-supported sub-populations like small businesses, indigenous populations, and rural communities?  Within this context, it is important for statistical agency decision-makers and staff to view their day-to-day work (the \textit{statistical purposes} they are pursuing) with not just respondent confidentiality in mind, but also with a strong eye towards the public benefit, objectivity, integrity, and representativeness of the statistics they are producing.\footnote{For some general discussion of these topics and related technical, managerial, and policy issues, see \cite{NAP27934} and \cite{FCSM2020DataQualityFramework}.}

\section{Statistical Purpose in the Statutory, Scientific, and Ethical Contexts}
\label{sec:frameworks}

Though the definition of \textit{statistical purpose} in the original 2002 CIPSEA does not explicitly include the additional notions of public benefit, objectivity, integrity, and representativeness, the 2019 revisions to CIPSEA enshrined in the Evidence Act do put the term squarely within this broader context. In fact, one of the most notable components of the Evidence Act was its codification of the four principal responsibilities of a federal statistical agency, namely to: ``(A) produce and disseminate relevant and timely statistical information; (B) conduct credible and accurate statistical activities; (C) conduct objective statistical activities; and (D) protect the trust of information providers by ensuring the confidentiality and exclusive statistical use of their responses'' (44 USC \S 3563(a)).\footnote{Prior to the Evidence Act, these responsibilities were originally enumerated in the 2014 Statistical Policy Directive \#1 ``Fundamental Responsibilities of Federal Statistical Agencies and Recognized Statistical Units'' issued by the Office of Management and Budget.}  

When interpreting the Evidence Act passage quoted above, it is useful to review the four related definitions provided in the Act:

\begin{quote}
    “(1) ACCURATE.—The term ‘accurate’, when used with respect to statistical activities, means statistics that consistently match the events and trends being measured. \\
    (2) CONFIDENTIALITY.—The term ‘confidentiality’ means a quality or condition accorded to information as an obligation not to disclose that information to an unauthorized party. \\
    (3) OBJECTIVE.—The term ‘objective’, when used with respect to statistical activities, means accurate, clear, complete, and unbiased. \\
    (4) RELEVANT.—The term ‘relevant’, when used with respect to statistical information, means processes, activities, and other such matters likely to be useful to policymakers and public and private sector data users.” (44 U.S.C. § 3563(d))
\end{quote}

Among responsibilities (A)-(D) above, \textit{protecting trust} and \textit{ensuring confidentiality} are most commonly associated with \textit{statistical purpose} as it has been used for most of the last century. \textit{Relevance} relates directly to the notions of public benefit and representativeness discussed above, ensuring that the statistics produced are germane to their intended uses and that they capture a complete enough picture of the nation to support necessary policymaking and decision-making. \textit{Objectivity}, in this context, relates to and extends the notion of \textit{functional separation} that was discussed above: are the agencies' statistical activities insulated from the administrative and enforcement functions of government and from political or partisan influence. This latter point does not mean that \textit{statistical purpose} requirements prohibit an NSO from producing objective, high-quality statistical information related to topics that may be at the core of policy, political, or partisan debates. Rather, this objectivity means that the collection, processing, and production of the statistical product is insulated from partisan influence that could undermine trust in the credibility or accuracy of the statistic. This responsibility for objectivity further obligates statistical agencies to avoid partisan interpretation of statistics--that is, a statistical agency can say `the statistic is X' but should not opine on whether that particular value of X is unacceptably low or high.\footnote{\cite{NAP27934} provides additional discussion of prohibitions against partisan interference.} Taken together, the fundamental responsibilities relating to \textit{relevance}, \textit{objectivity}, and \textit{confidentiality} are fully consistent with the various explicit and implicit aspects of \textit{statistical purpose} that we explored above. 

Understanding the role and importance of \textit{credibility} and \textit{accuracy} in the context of \textit{statistical purpose} is not something for which the historical context can provide much in the way of guidance. Instead, it is instructive to turn to two additional frameworks on which scientists and statisticians routinely rely: the \textit{responsible conduct of science} and the \textit{professional practice of statistics}. 

Statistical agencies are, at their core, scientific organizations, and as the thirteen principal U.S. statistical agencies affirmed in a 2012 joint statement, ``scientific methods play a critical role in maximizing the quality, objectivity, and credibility of information collected and disseminated by the principal statistical agencies'' \citep{fss}. Because of this, the statistical agencies place great emphasis on adhering to the principles of scientific integrity. This requires a firm commitment to protecting the integrity of the scientific process itself, but also a commitment to ensuring the transparency of that process and the accountability of those involved.   

To understand the meaning and requirements of scientific rigor in the context of the statistical activities performs, it is helpful to turn to established, generally accepted ethical frameworks for statistics as a profession, including those of the American Statistical Association (ASA) \citep{asa}  or the International Statistical Institute \citep{isi}. While there is substantial consistency between these professional ethical frameworks and the general principles of scientific integrity, the professional frameworks provide greater specificity on issues and concerns relating to statistical methods and analysis. The ASA's Ethical Guidelines for Statistical Practice, for example, includes specific principles relating to data integrity, including guidelines regarding transparency about the data generation and collection process, known sources of error, data processing and transformation procedures, and how missing data are handled, among many others \citep[Principle B]{asa}.

Adherence to these scientific integrity principles and ethical guidelines in real world scenarios does often require a degree of subjective reflection on how they would apply in specific circumstances. The level of abstraction and generality in principles and guidelines, though informative and applicable to a broad range of situations and contexts, does not provide specific and unambiguous standards and procedures for agency staff to follow as they perform their day-to-day work supporting the agencies' \textit{statistical purposes}. To achieve that level of specificity, and to ensure methodological consistency across the agency's portfolio of statistical products (a necessary prerequisite for maintaining the external \textit{credibility} of those products), agencies define and adopt statistical quality standards that their products must adhere to. These standards, which are often extensive and span the entire statistical product lifecycle from initial survey planning through to final product dissemination of results and analysis \citep{sqs}, are intended to translate principles of scientific integrity and ethical statistical practice into clear and definitive metrics and processes for agency staff to follow. 

Lastly, it is important to acknowledge that statistical agencies exist in an environment of constrained (sometimes substantially so) resources. Efforts to maximize the relevance, timeliness, representativeness, confidentiality, credibility, or accuracy of a statistical product are not costless, and tradeoffs inevitably have to be made. Extending field collection, for example, may help to increase response rates, which in turn may improve accuracy and representativeness, but that extension can negatively impact timeliness (if it delays the scheduled release of the statistic) and can increase costs (which may preclude other necessary or advisable statistical activities). Consequently, these objectives, as important and central as they are to the responsibilities of a statistical agency in achieving their \textit{statistical purposes}, cannot always be unwaveringly maximized. Compromises will often have to be made in the broader pursuit of the agency's mission to produce high quality, credible statistics.  The frameworks discussed above (scientific integrity principles, professional ethical guidelines, and statistical quality standards) can help agency decision-makers navigate these compromises, as can robust engagement with internal and external stakeholders. Better knowledge of these stakeholders' needs (or concerns) regarding these statistics or the purposes for which they will be used, can help prioritize among these competing objectives. Agencies' legal obligations and responsibilities, interpreted through the lens of their commitment to scientific integrity and the ethical practice of statistics, and informed by the needs and perspectives of data subjects and data users, together define and illuminate what is, or is not, an acceptable and responsible \textit{statistical purpose} for an agency to pursue. \cite{eltinge2025some, eltinge2026ethical} provides additional discussion of related topics.

\section{The International Perspective}
\label{sec:international}

Though the discussion above is focused on the U.S. statistical agency experience, with a particular focus on the experience of the Census Bureau over the past century, similar terminology is used by NSOs internationally with comparable meanings and ambiguity. EUROSTAT, for example, defines \textit{statistical purposes} in terms of ``the development and production of statistical results and analyses'' \citep{eurostat}. Statistics Canada uses a similar definition, the ``description or analysis of characteristics of a population to which the individual belongs'' \citep{statcan}. The United Nations Statistical Division uses an even more ambiguous definition, namely ``tasks aimed at developing, producing, and disseminating official statistics, including experimenting and testing'' \citep{unsd}. Both the UK Office of National Statistics and Statistics New Zealand make explicit reference, as President Taft's proclamation did, to the public benefit component of statistical purposes, stating ``the production of official statistics that serve the public good'' \citep{ukons} and ``[the] Statistician must be satisfied research is in public interest'' \citep{NewZealand2022DataStatisticsActS49}. Though this paper largely focuses on the U.S. federal statistical experience, this broad similarly in notions of \textit{statistical purposes} internationally suggests that the conclusions may have broader applicability to NSOs worldwide.

\section{The Case for a Broader Definition}
\label{sec:broader}
The preceding sections show how the concept of statistical purposes provides a notional set of conceptual guardrails for statistical agencies that go well beyond the term's commonly used explicit statutory definition and implicitly incorporating much from the scientific and professional frameworks discussed above. The logical next step would be to develop a broader definition of statistical purpose that serves to make the implicit meaning explicit. For that, it is helpful to consider the relationship between statistical purpose and the data subject within the context of the seminal ethical frameworks relating to information collection and the individual: the Fair Information Practice Principles in 1973, the Belmont Report in 1979, and the Menlo Report in 2012.

A 1966 initiative to improve the efficiency of the federal statistical system led to a short-lived proposal to create a ``National Data Bank''\citep{janet1995norwood, kaysen1969report}. Opposition to the proposal on privacy grounds was swift and led to a broader awareness of the growing privacy concerns relating to the rapid rise of automated data systems within the federal government. This led to the creation, in 1972, of the Department of Health, Education, and Welfare (HEW) Secretary's Advisory Committee on Automated Personal Data Systems. That committee's final report ``Records, Computers, and the Rights of Citizens'' established the federal ``Code of Fair Information Practice'' (also known as the Fair Information Practice Principles, or FIPPs) to safeguard data subjects from ``arbitrary or abusive'' uses of their information in an increasingly digital world \citep{HEW1973FairInformationPractices}.\footnote{Note that the use of the term ``fair'' in the Hew Report is distinct from the usage of the acronym FAIR (Findable, Accessible, Interoperable, Reusable) intended to describe principles for scientific data management and stewardship as discussed in \cite{wilkinson2016fair}.} These safeguards included the need for proper notice about what information is being collected and how it is being used, limitations on the re-use of data for purposes other than those for which it was collected, the ability to correct information about oneself, and the responsibility for agencies to prevent the misuse of those data. These principles served as the cornerstone of subsequent federal privacy law, including the Privacy Act of 1974\footnote{Purpose specification and limitation becomes especially challenging in the context of statistical agencies re-use of data originally collected for administrative purposes, such as tax or Social Security information. In these cases, the Privacy Act provides a mechanism for providing notice about these secondary statistical uses to data subjects through the ``Routine Uses'' section of agencies' System of Records Notices. For more information, see Section 7 of Office of Management and Budget Memorandum M-14-06 ``Guidance for Providing and Using Administrative Data for Statistical Purposes''\citep{omb2014m1406}}.  More importantly they have informed the broader notion of what privacy and respect for data subjects mean in the information era, and how transparency and responsible and constrained use of data are essential to meeting those obligations and protecting individuals from being unnecessarily or inadvertantly harmed by government uses of their data.

A few years later, in the aftermath of the egregious ethical abuses of the Tuskegee Syphilis Study, the National Commission for the Protection of Human Subjects of Biomedical and Behavioral Research published the Belmont Report \citep{NationalCommission1979BelmontReport}.  This report identified three basic ethical principles that should guide any research involving human subjects. The first, ``Respect for Persons'' requires that individuals be treated as autonomous agents and that their judgment (consent) should be respected whenever possible, and with full information given to make an informed decision. The second principle of the Belmont Report, ``Beneficence'', establishes an expectation of avoiding intentional harm to the subject and of maximizing potential benefits while minimizing potential harms. The third principle, ``Justice'', establishes the ethical expectation that the benefits and burdens of the research should be fairly distributed. 

In 2012, the Department of Homeland Security's ``Menlo Report'' built on and extended the Belmont Report's principles into the domain of information and communication technology, adding ethical expectations for considering stakeholder perspectives (especially the potential for group harm) and for respecting the Public Interest, especially with regard to transparency and accountability. To this latter point, the report states ``Transparency entails clearly communicating the purposes of research---why data collection...is required to fulfill those purposes...It also involves clear communication of risk assessment and harm minimization...'' The report goes on to state ``Accountability demands that research methodology, ethical evaluations, data collected, and results generated should be documented and made available responsibly in accordance with balancing risks and benefits. Data should be available for legitimate research, policy-making, or public knowledge...'' \citep{dittrich2012menlo}.  

Focusing as they do on the ethics of human subjects research, the ethical frameworks of the Belmont and Menlo reports do not directly translate to the broader work of statistical agencies. Taken together with the FIPPs, however, these frameworks can provide important insight into what the responsible and appropriate use and  stewardship of data subjects' information actually means and how it is critical to promoting and preserving the public's trust in the agency's statistics and statistical activities. 

From the FIPPs, one sees the importance of transparency about the statistical agency's information collections, the purposes for which those data are being collected, and how they will be safeguarded against misuse. From the Belmont Report, we see the importance of preventing intentional harm to the individual from the use of their data, along with the goal of maximizing the potential benefits that derive from the use of those data, while minimizing the potential risks. The Belmont Report also stresses the importance of fairly distributing, across society, the burden and benefits of the information's collection and use. This requires both the responsible minimization of data collection (only collecting the information you need for the intended purpose) and ensuring that the statistics being produced provide a broader societal benefit, rather than merely serving narrower private, parochial, or partisan interests. This focus on the importance of public benefit is also echoed by the Menlo Report, with its focus on the importance of data being available for ``\textit{legitimate} research, policy-making, or public knowledge'' (emphasis added). Reflecting on the Census Bureau's experience during World War II (discussed in section \ref{sec:origins}), the Menlo Report's inclusion of the need to avoid intentional group harm is also highly relevant. Lastly, the Menlo Report's emphasis on accountability, particularly with regard to methodology, serves as an important reminder that ``statistics is the study of uncertainty,'' \citep{https://doi.org/10.1111/1467-9884.00238} and that the ethical practice of statistics requires transparency about known limitations of the statistics being produced\footnote{See \cite{asa} principle B2} and, to the maximum extent practicable, the quantification of the inherent uncertainty associated with those statistics.

Synthesizing these objectives within the context of existing explicit definitions and implicit interpretations of ``statistical purposes,'' we can envision a broader, applied definition of the term that can better guide and inform statistical agency decision-making as they seek to conduct their statistical activities while being good, responsible, and above all \textit{trusted} stewards of the public's information. To that end, we propose the following as a more informative and helpful definition for statistical agencies to consider:
\begin{quote}
\textit{``Statistical purposes'' entail the activities of an NSO in support of the production and dissemination of statistics for the public benefit that are accurate, relevant, and objective. Statistical purpose requires a firm commitment to scientific integrity at all stages of the information lifecycle with robust transparency about the statistical activities being performed and about the known limitations of the statistics being produced. In performing these activities, NSOs must acknowledge that they are acting as stewards of the public's information, and must maintain the public's trust that the identifiable information provided to the agency will remain confidential and will not be used with the intent to directly harm or otherwise adversely impact any individual data subject. Statistical purposes also means avoiding, where reasonably practicable, the production or dissemination of statistical information or other statistical activities that are intended to harm groups of data subjects, that lack a clearly defined public benefit, or that are intended to serve purely private, parochial, or partisan objectives.}
\end{quote}
While the preceding definition specifically references NSOs, it may also have relevance for academic or private sector organizations with similar missions.

\section{Conclusion and Suggestions for Future Research}
\label{sec:conclusion}

This paper has explored the fundamental concept of “statistical purposes only” that NSOs routinely use in communication with prospective data subjects, data users and other stakeholders.  A review of applicable U.S. law, and related policy documents, has noted the importance of two complementary criteria. Firstly, that the provided information will only be used for the production of statistical information about relatively large population aggregates, with the intention of creating a general public benefit. And secondly, that the NSO will properly safeguard the confidentiality of the information provided by or about data subjects and that such information will not be used for legal or regulatory action against those individuals or organizations or in other ways that could cause them individual harm. In addition, we have noted important features of the societal and scientific context within which NSOs apply these concepts.  Those features include indications of declining public trust in government and related institutions, a general reduction in survey response rates, and expressions of increasing public interest in the availability of statistical information for evidence-based decision-making and for use by the private sector and academia.  

Practical navigation of those contextual factors requires careful attention to fundamental principles of sound scientific processes and statistical ethics.  To that end, we have proposed a broader, applied definition of ``statistical purposes'' that we hope statistical agencies can consider in the context of their decision-making about appropriate and responsible data use and their role as stewards of the public's information.

An important area for methodological and empirical research is how prospective survey respondents and other stakeholders understand and interpret the term “statistical purposes only” and how those interpretations affect trust, participation, and engagement. Such research could also assess how those interpretations may vary by demographic characteristics and context, as well as by their roles as survey respondents, administrative program participants, data users, intermediaries, or policymakers. Research in this area could also evaluate how members of these groups respond to different written or spoken explanations of “statistical purposes only” in specific context, extending prior work such as Landreth et al. (2008) and Hunter-Childs et al. (2019), and helping to inform future communication and engagement strategies with each of these stakeholder segments.

\section{Disclosure statement}\label{disclosure-statement}

 No rights reserved. This work was authored as part of the authors' official duties as Employees of the United States Government and is therefore a work of the United States Government. In accordance with 17 U.S.C. 105, no copyright protection is available for such works under U.S. law.

  \bibliography{bibliography.bib}

@misc{USCB:privacyprinciples,
  author = {{U. S. Census Bureau}},
  title = {Our Privacy Principles},
  year = 2006,
  url = {https://www.census.gov/about/policies/privacy/data_stewardship/our_privacy_principles.html},
  urldate = {2025-02-03}
}

@misc{BLS:mxp,
  author = {{U.S. Bureau of Labor Statistics}},
  title = {OImport/Export Price Indexes (MXP) Survey Respondents: Confidentiality of Data Collected by BLS for Statistical Purposes},
  year = 2019,
  url = {https://www.bls.gov/respondents/mxp/confidentiality.htm},
  urldate = {2019-05-17}
}

@misc{NCHS:NEHRS,
  author = {{National Center for Health Statistics}},
  title = {National Electronic Health Records Survey: NEHRS Participants},
  year = 2024,
  url = {https://www.cdc.gov/nchs/nehrs/partcipant/},
  urldate = {2024-12-13}
}

@article{landreth2008report,
  title={Report of Cognitive Testing of Privacy and Confidentiality-Related Statements in Respondent Materials for the 2010 Decennial: Results from Cognitive Interview Pretesting with Volunteer Respondents.” US Census Bureau},
  author={Landreth, Ashley and Gerber, Eleanor and DeMaio, Theresa},
  journal={Statistical Research Division Report Series (Survey Methodology\# 2008-4). On-line, available: http://www. census. gov/srd/papers/pdf/rsm2008-04. pdf},
  volume={5},
  year={2008}
}

@misc{NCES:standards,
  author = {{National Center for Education Statistics}},
  title = {Statistical Standard 4-2: Maintaining Confidentiality},
  year = 2002,
  url = {https://nces.ed.gov/statprog/2002/std4_2.asp},
  urldate = {2025-02-21}
}

@misc{OMB:stats,
  author = {{Office of Management and Budget}},
  title = {The Federal Statistical System},
  year = 2025,
  url = {https://www.statspolicy.gov/about/#statistical-agencies},
  urldate = {2025-02-21}
}

@article{reiss2021public,
  title={Public goods},
  author={Reiss, Julian},
  journal={Stanford Encyclopedia of Philosophy},
  year={2021},
  url={https://plato.stanford.edu/entries/public-goods/?fbclid=IwAR3F2UdGd6nYoxsDdeTcyNtNNYHBOzqtLxPHTrzIZRR3C0XuWBSaAquCEag},
}

@article{oed,
  author = {{Oxford University Press}},
  title = {Statistics},
  journal = {Oxford English Dictionary},
  year = 2024,
  url = {https://doi.org/10.1093/OED/2506503088},
  urldate = {2025-02-21}
}

@online{taft,
  author = {Taft, William H.},
  title = {A Proclamation},
  year = 1910,
  url = {https://www2.census.gov/about/history/other-resources-reference/artifacts/1910-census-proclamation.jpg},
  urldate = {2025-02-21}
}

@article{gatewood2001monograph,
  title={A monograph on confidentiality and privacy in the US Census},
  author={Gatewood, George and Micarelli, WF},
  journal={US Bureau of the Census, Washington},
  year={2001}
}

@inproceedings{seltzer2007census,
  title={Census confidentiality under the second war powers act (1942-1947)},
  author={Seltzer, William and Anderson, Margo},
  booktitle={Presentation at the Population Association of America Annual Meeting},
  year={2007}
}

@online{events,
  author = {{U.S. Census Bureau}},
  title = {Events in the Chronological Development of Privacy and Confidentiality at the U.S. Census Bureau},
  year = 2002,
  url = {https://www2.census.gov/about/history/agency-history/privacy-confidentiality-chronological-development.pdf},
  urldate = {2025-02-24}
}

@book{united1977personal,
  title={Personal Privacy in an Information Society: The Report of the Privacy Protection Study Commission},
  author={{Privacy Protection Study Commission}},
  volume={2},
  year={1977},
  publisher={The Commission}
}

@online{pew,
  author = {{Pew Research Center}},
  title = {Public Trust in Government: 1958-2024},
  year = 2025,
  url = {https://www.pewresearch.org/politics/2024/06/24/public-trust-in-government-1958-2024/},
  urldate = {2025-02-24}
}

@article{hunter2019trust,
  title={Trust and credibility in the US Federal Statistical System},
  author={Hunter Childs, Jennifer and Clark Fobia, Aleia and King, Ryan and Morales, Gerson},
  journal={Survey Methods: Insights from the Field},
  pages={1--10},
  year={2019},
  publisher={DEU}
}

@article{czajka2016background,
  title={Background paper declining response rates in federal surveys: Trends and implications},
  author={Czajka, John L and Beyler, Amy},
  journal={Mathematica policy research},
  volume={1},
  pages={1--86},
  year={2016}
}

@online{acs,
  author = {{U.S. Census Bureau}},
  title = {American Community Survey Response Rates},
  year = 2024,
  url = {https://www.pewresearch.org/politics/2024/06/24/public-trust-in-government-1958-2024/},
  urldate = {2025-02-24}
}

@article{madden2015americans,
  title={Americans’ attitudes about privacy, security and surveillance},
  author={Madden, Mary and Rainie, Lee},
  year={2015},
  publisher={Pew Research Center}
}

@article{singer1992confidentiality,
  title={Confidentiality assurances in surveys: Reassurance or threat?},
  author={Singer, Eleanor and Hippler, Hans-Juergen and Schwarz, Norbert},
  journal={International journal of Public Opinion research},
  volume={4},
  number={3},
  pages={256--268},
  year={1992},
  publisher={Oxford University Press}
}

@book{landreth2001sipp,
  title={SIPP Advance Letter Research: Cognitive Interv [i] ew Results, Implications, \& Letter Recommendations},
  author={Landreth, Ashley},
  year={2001},
  publisher={US Bureau of the Census}
}

@article{gerber2000privacy,
  title={Privacy Schemas and Data Collection: An Ethnographic Account},
  author={Gerber, Eleanor},
  journal={Final Report, Census},
  year={2000},
  publisher={Citeseer}
}

@online{fss,
  author = {{U.S. Federal Statistical System}},
  title = {Statement of Commitment to Scientific Integrity by Principle Statistical Agencies},
  year = 2012,
  url = {https://www.ers.usda.gov/publications/pub-details?pubid=42788},
  urldate = {2025-02-25}
}

@online{integrity,
  author = {{U.S. Census Bureau}},
  title = {Census Bureau Scientific Integrity Policy},
  year = 2024,
  url = {},
  urldate = {}
}

@online{asa,
  author = {{American Statistical Association}},
  title = {Ethical Guidelines for Statistical Practice},
  year = 2022,
  url = {https://www.amstat.org/your-career/ethical-guidelines-for-statistical-practice},
  urldate = {2025-02-25}
}

@online{isi,
  author = {{International Statistical Institute}},
  title = {Declaration on Professional Ethics},
  year = 2010,
  url = {https://isi-web.org/declaration-professional-ethics},
  urldate = {2025-02-25}
}

@online{sqs,
  author = {{U.S. Census Bureau}},
  title = {Statistical Quality Standards},
  year = 2023,
  url = {https://www2.census.gov/about/policies/quality/quality-standards.pdf},
  urldate = {2025-02-25}
}

@online{eurostat,
  author = {{European Commission, Eurostat}},
  title = {Statistical requirements compendium},
  year = 2023,
  url = {https://data.europa.eu/doi/10.2785/562159},
  urldate = {2025-02-25}
}

@online{statcan,
  author = {{Statistics Canada}},
  title = {Statistics Canada Quality Assurance Framework},
  year = 2017,
  url = {https://www150.statcan.gc.ca/n1/en/pub/12-586-x/12-586-x2017001-eng.pdf?st=9tr3nuNz},
  urldate = {2023-05-22}
}

@online{unsd,
  author = {{United Nations Statistics Division}},
  title = {United Nations National Quality Assurance Frameworks Manual for Official Statistics},
  year = 2019,
  url = {https://unstats.un.org/unsd/dnss/docs-nqaf/UN_NQAF_Manual-Unedited_manuscript_of_3_May_2019.pdf},
  urldate = {}
}

@online{ukons,
  author = {{United Kingdom Office of National Statistics}},
  title = {Data Protection: Privacy Information},
  year = 2023,
  url = {ttps://www.ons.gov.uk/aboutus/transparencyandgovernance/dataprotection},
  urldate = {}
}

@article{nickson2024statistics,
  title={Statistics for the public good: What it means and why it matters},
  author={Nickson, Sofi},
  journal={Statistical Journal of the IAOS},
  volume={40},
  number={3},
  pages={501--509},
  year={2024},
  publisher={SAGE Publications Sage UK: London, England}
}

@BOOK{NAP27934,
  author    = {{National Academies of Sciences, Engineering, and Medicine}},
  editor    = {Melissa Chiu and Jennifer Park},
  title     = {Principles and Practices for a Federal Statistical Agency: Eighth Edition},
  isbn      = {978-0-309-72543-9},
  doi       = {10.17226/27934},
  url       = {https://nap.nationalacademies.org/catalog/27934/principles-and-practices-for-a-federal-statistical-agency-eighth-edition},
  year      = 2025,
  publisher = {The National Academies Press},
  address   = {Washington, DC}
}

@techreport{FCSM2020DataQualityFramework,
  author      = {{Federal Committee on Statistical Methodology}},
  title       = {{A Framework for Data Quality}},
  institution = {{Federal Committee on Statistical Methodology}},
  number      = {FCSM-20-04},
  year        = {2020},
  month       = sep,
  url         = {https://statspolicy.gov/assets/fcsm/files/docs/FCSM.20.04_A_Framework_for_Data_Quality.pdf}
}

@article{eltinge2025some,
  title={Some Dimensions of Statistical Ethics and Scientific Integrity That Warrant Exploration Through Empirical Studies of Stakeholder Information Needs},
  author={Eltinge, John L},
  journal={Journal of Official Statistics},
  volume={41},
  number={3},
  pages={804--812},
  year={2025},
  publisher={SAGE Publications Sage UK: London, England}
}

@article{eltinge2026ethical,
  title={Ethical Issues in the Dynamic Environment for Public-Stewardship Statistical Information},
  author={Eltinge, John L},
  journal={Data Science in Science},
  volume={5},
  number={1},
  pages={2605771},
  year={2026},
  publisher={Taylor \& Francis}
}

@misc{NewZealand2022DataStatisticsActS49,
  author       = {{New Zealand}},
  title        = {{Data and Statistics Act 2022, s 49}},
  year         = {2022},
  howpublished = {\url{https://www.nzlii.org/nz/legis/consol_act/dasa2022188/s49.html}},
  note         = {Section 49, Research is in public interest; accessed May 1, 2026}
}

@techreport{VillaRoss2023FederalFundsFY2021,
  author      = {Villa Ross, Ceci A.},
  title       = {{Uses of Decennial Census Programs Data in Federal Funds Distribution: Fiscal Year 2021}},
  institution = {{U.S. Census Bureau}},
  year        = {2023},
  month       = jun,
  type        = {Working Paper},
  url         = {https://www2.census.gov/library/working-papers/2023/decennial/census-data-federal-funds-fy-2021.pdf}
}

@techreport{NationalCommission1979BelmontReport,
  author      = {{National Commission for the Protection of Human Subjects of Biomedical and Behavioral Research}},
  title       = {{The Belmont Report: Ethical Principles and Guidelines for the Protection of Human Subjects of Research}},
  institution = {{U.S. Department of Health, Education, and Welfare}},
  year        = {1979},
  month       = apr,
  day         = {18},
  url         = {https://www.hhs.gov/ohrp/regulations-and-policy/belmont-report/read-the-belmont-report/index.html}
}

@techreport{HEW1973FairInformationPractices,
  author      = {{U.S. Department of Health, Education, and Welfare, Secretary's Advisory Committee on Automated Personal Data Systems}},
  title       = {{Records, Computers, and the Rights of Citizens: Report of the Secretary's Advisory Committee on Automated Personal Data Systems}},
  institution = {{U.S. Department of Health, Education, and Welfare}},
  year        = {1973},
  month       = jul,
  url         = {https://epic.org/documents/hew1973report/},
  note        = {Foundational source for the Fair Information Practice Principles}
}

@misc{janet1995norwood,
  title={Norwood. Organizing to Count: Change in the Federal Statistical System},
  author={Norwood, Janet L},
  year={1995},
  publisher={Urban Institute Press}
}

@article{kaysen1969report,
  title={Report of the Task Force on the Storage of and Access to Government Statistics},
  author={Kaysen, Carl and Holt, Charles C and Holton, Richard and Kozmetsky, George and Morrison, H Russell and Ruggles, Richard},
  journal={The American Statistician},
  volume={23},
  number={3},
  pages={11--19},
  year={1969},
  publisher={Taylor \& Francis}
}

@techreport{dittrich2012menlo,
  title={The Menlo Report: Ethical principles guiding information and communication technology research},
  author={Dittrich, David and Kenneally, Erin and others},
  year={2012},
  institution={US Department of Homeland Security}
}

@article{https://doi.org/10.1111/1467-9884.00238,
author = {Lindley, Dennis V.},
title = {The Philosophy of Statistics},
journal = {Journal of the Royal Statistical Society: Series D (The Statistician)},
volume = {49},
number = {3},
pages = {293-337},
doi = {https://doi.org/10.1111/1467-9884.00238},
url = {https://rss.onlinelibrary.wiley.com/doi/abs/10.1111/1467-9884.00238},
eprint = {https://rss.onlinelibrary.wiley.com/doi/pdf/10.1111/1467-9884.00238},
year = {2000}
}

@misc{omb2014m1406,
  author       = {{Office of Management and Budget}},
  title        = {{Guidance for Providing and Using Administrative Data for Statistical Purposes}},
  howpublished = {Memorandum M-14-06},
  institution  = {{Executive Office of the President, Office of Management and Budget}},
  year         = {2014},
  month        = feb,
  day          = {14},
  url          = {https://www.whitehouse.gov/wp-content/uploads/legacy_drupal_files/omb/memoranda/2014/m-14-06.pdf},
  note         = {Memorandum for the heads of executive departments and agencies, from Sylvia M. Burwell, Director}
}

@article{wilkinson2016fair,
  title={The FAIR Guiding Principles for scientific data management and stewardship},
  author={Wilkinson, Mark D and Dumontier, Michel and Aalbersberg, IJsbrand Jan and Appleton, Gabrielle and Axton, Myles and Baak, Arie and Blomberg, Niklas and Boiten, Jan-Willem and da Silva Santos, Luiz Bonino and Bourne, Philip E and others},
  journal={Scientific data},
  volume={3},
  number={1},
  pages={1--9},
  year={2016},
  publisher={Nature Publishing Group}
}

\end{document}